\documentclass[aip,preprint,jap]{revtex4-1}

\usepackage{graphicx} 
\usepackage{amsmath}
\usepackage{amssymb}
\usepackage{natbib}

\begin{document}

\bibliographystyle{aipnum4-1}

\title{Infrared signatures of high carrier densities induced in semiconducting poly(3-hexylthiophene) by fluorinated organosilane molecules}
\date{\today}

\author{O. Khatib}
\email[Electronic address: ]{okhatib@physics.ucsd.edu}
\affiliation{Department of Physics, University of California, San Diego, La Jolla, California 92093, USA}
\author{B. Lee}
\affiliation{Department of Physics and Astronomy, Rutgers University, Piscataway, New Jersey 08854, USA}
\author{J. D. Yuen}
\affiliation{Center for Polymers and Organic Solids, University of California, Santa Barbara, Santa Barbara, California 93106, USA}
\author{Z. Q. Li}
\affiliation{Department of Physics, Columbia University, New York, New York 10027, USA} 
\author{M. Di Ventra}
\affiliation{Department of Physics, University of California, San Diego, La Jolla, California 92093, USA} 
\author{A. J. Heeger}
\affiliation{Center for Polymers and Organic Solids, University of California, Santa Barbara, Santa Barbara, California 93106, USA}
\author{V. Podzorov}
\affiliation{Department of Physics and Astronomy, Rutgers University, Piscataway, New Jersey 08854, USA}
\author{D. N. Basov}
\affiliation{Department of Physics, University of California, San Diego, La Jolla, California 92093, USA}  

\begin{abstract}
We report on infrared (IR) absorption and dc electrical measurements of thin films of poly(3-hexylthiophene) (P3HT) that have been modified by a fluoroalkyl trichlorosilane (FTS). Spectra for FTS-treated films were compared to data for electrostatically-doped P3HT in an organic field-effect transistor (OFET). The appearance of a prominent polaron band in mid-IR absorption data for FTS-treated P3HT supports the assertion of hole doping via a charge-transfer process between FTS molecules and P3HT.  In highly-doped films with a significantly enhanced polaron band, we find a monotonic Drude-like absorption in the far-IR, signifying delocalized states.  Utilizing a simple capacitor model of an OFET, we extracted a carrier density for FTS-treated P3HT from the spectroscopic data.  With carrier densities reaching 10$^{14}$ holes/cm$^2$, our results demonstrate that FTS doping provides a unique way to study the metal-insulator transition in polythiophenes.
\end{abstract}

\maketitle

\section{Introduction}

Conjugated polymers, and specifically polythiophenes, are attractive materials for low-cost, large area, flexible electronics applications because of their solution processability, superior film-forming properties, and comparatively high mobilities.\cite{heeger-rev-rmp2001,mccullough-am2006,frisbie-peo-afm2006,yang-apl2007} Because these systems are insulators with a moderate band gap, it is necessary to introduce mobile charges through doping or electric field gating, in order to initiate electrical transport.  The quest to achieve a comprehensive understanding of charge carrier dynamics in the polymer host also requires extending the carrier densities toward the boundary of the insulator-to-metal transition (IMT).  The IMT region is of fundamental interest as well, as some of the most fascinating electronic and magnetic many-body effects occur when both organic and inorganic insulators are driven towards the metallic state by photoexcitation, gating, or doping.\cite{ahn-nature2003,ahn-rmp2006,morpurgo-prl2004}

To go beyond the charge density limits of conventional organic field-effect transistors (OFETs) with oxide insulators (roughly 10$^{13}$/cm$^{2}$ ), alternative methods have been introduced, including OFETs employing polymer electrolyte gating,\cite{dhoot-pnas2006,yuen-jacs2007,panzer-jacs2007,herlogsson-am2008,zhu-apa2009} and the use of fluorinated organosilane molecules (known to self-assemble on surfaces) to chemically oxidize and thereby dope the organic semiconductor.\cite{kao-afm2009,podzorov-natmat2007} In this work, we explore the latter approach by modifying the electronic properties of the polymer poly(3-hexylthiophene) (P3HT) using fluorinated organosilanes. The organosilane molecules incorporate into the polythiophene structure, hydrolyze, and partially cross-link, forming a network that induces a strong p-type doping of P3HT.\cite{kao-afm2009} The particular organosilane used in this work is (Tridecafluoro-1,1,2,2-tetrahydrooctyl)trichlorosilane (C$_{8}$H$_{4}$F$_{13}$SiCl$_3$), or simply fluoroalkyl trichlorosilane (FTS).   
  
We have utilized infrared (IR) spectroscopy:  a tool ideally suited for investigating fundamental electronic processes in organic semiconductors.\cite{pope} IR measurements directly probe electronic excitations associated with injected charges, and are complementary to transport measurements. With IR spectroscopy, we are able to characterize the low-energy excitations induced in FTS-modified structures by comparing the IR absorption of FTS-treated films to that of the same polymer film electrostatically doped in an OFET.  At the highest carrier densities, we have observed both Drude absorption (indicative of delocalized states in a metal) and a broad mid-IR absorption (indicative of self-localized polaron states).  We therefore demonstrate that FTS modification is a useful technique for exploring and exploiting the onset of metallic transport in polymers.

\section{Experimental}

\begin{figure}
\centering
   \includegraphics[width=3.375in, height=2.53in]{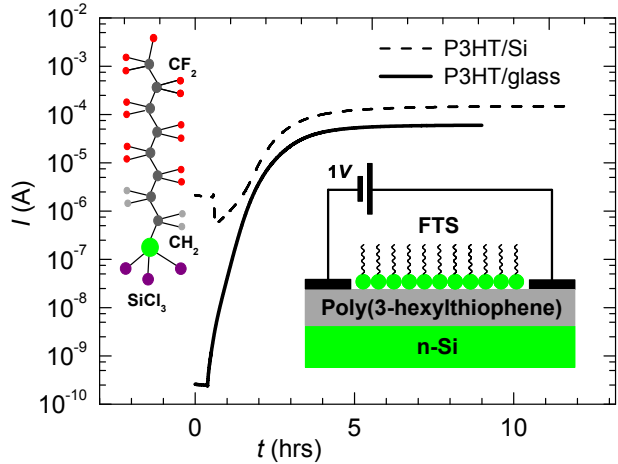}
     \caption{DC current flowing through a 10-15 nm-thick P3HT film exposed to the fumes of (trideca�uoro-1,1,2,2-tetrahydrooctyl)trichlorosilane (FTS ) as a function of exposure time.  Dashed line: P3HT on a semi insulating (transparent for IR) Si(111) substrate. Solid line: P3HT on a glass substrate. Measurements are performed with V$ = 1 V$ applied between graphite contacts that define a 1 x 1 cm$^{2}$ P3HT film. The left inset shows the structure of FTS molecules. The right inset shows a conceptual geometry of 2-probe samples. The apparent large background conductivity in the P3HT/Si device is due to the residual bulk conduction through the Si substrate.} 
     \label{DCplot}
\end{figure}

All IR and DC transport measurements reported here have been performed at room temperature using two types of devices: (a) two-terminal structures and (b) three terminal OFETs, shown in the insets of Figures \ref{DCplot} and \ref{OFET}, respectively.  Two-terminal devices were fabricated by spin-coating a 10 - 15 nm-thick P3HT film onto an intrinsic H-terminated Si(111) substrate with bulk resistivity 30 $k\Omega\cdot$ cm, or onto a glass slide.  Graphite contacts were painted on top of the polymer film, separated by ~ 0.5 - 1 cm (Fig. \ref{DCplot} inset). Three-terminal P3HT-based OFETs were prepared by depositing thin (4 - 6 nm) film of P3HT on SiO$_2$/n-Si substrates (inset in bottom panel of Fig. \ref{OFET}).  In the OFETs, 200 nm-thick SiO$_2$ on n-Si wafers were used, with the doping level suitable for IR transmission measurements.  Gold source and drain electrodes were patterned on top of SiO$_2$ and spaced roughly 200 microns apart as previously described in Ref. \onlinecite{zq-nanolett2006}.

IR measurements were performed ex-situ, i.e. after the P3HT samples have been brought to a highly conducting state by exposure to FTS. IR transmission measurements with a spectral resolution of 8 cm$^{-1}$ were carried out in the mid-IR range (400 - 7000 cm$^{-1}$) and were extended to the far-IR region (down to 40 cm$^{-1}$) for selected structures.   We employed a home-built IR microscope with the focus size reduced down to $d$ = 200 $\mu$m, allowing us to explore the uniformity of electronic properties along the sample surface.  Typically we recorded transmission spectra of an FTS-treated polymer, T$_{tr}(\omega)$, normalized by the transmission of a pristine sample, T$_{p}(\omega)$, on the same substrate.  It is customary to plot these data in the form of the change in absorption ($\Delta\alpha$d) defined as: 

\begin{equation}
\Delta\alpha d  = 1- T_{tr}(\omega)/T_{p}(\omega)).
\label{fts}
\end{equation}

For the OFETs we characterized transmission, T(V$_{gs})$, at various gate voltages, V$_{gs}$, and normalized these data by T(V$_{gs}=0 V)$.  OFET data are presented as:

\begin{equation}
\Delta\alpha d  = 1 - T(V_{gs})/T(V_{gs}=0V),
\label{ofet}
\end{equation}

where d is the thickness of the accumulation layer (typically 1 - 2 nm) and T(V$_{gs}=0 V)$ corresponds to the transmission at zero V$_{gs}$.

\section{Results}

Figure \ref{DCplot} shows DC transport data for a P3HT film obtained during exposure to FTS vapor.  A dramatic increase of the DC current, by 5 - 6 orders of magnitude, is observed within the first few hours of the treatment.  The large background conductivity of P3HT/Si(111) samples before the FTS exposure (dashed line) is due to the shunt by the 30 k$\Omega\cdot$cm substrate.  The magnitude of the current in the saturated regime is similar for both P3HT/Si(111) and P3HT/glass. The effect of FTS on conductivity is persistent:  FTS-altered P3HT shows no degradation of the high-conductivity state, provided the samples are stored in high vacuum or in an
atmosphere of dry non-polar gases.  The conductivity lasts for weeks in ambient atmosphere albeit with some gradual degradation accelerated by humidity.  FTS-induced modification of the DC conductivity is similar to results reported for FTS layers grown at the surface of organic molecular crystals.\cite{podzorov-natmat2007} For molecular crystals, the FTS is restricted to the surface of the crystal; since FTS molecules cannot penetrate the tightly packed molecular structure, they form a self-assembled monolayer at the surface.\cite{podzorov-natmat2007} In a polymer with weak interchain interactions, however, the entire volume of a 10 nm film reacts with FTS, evidenced by a change in the color of P3HT from purple to transparent as the result of FTS exposure.\cite{kao-afm2009}

\begin{figure}
\centering
   \includegraphics[width=3.375in, height=2.53in]{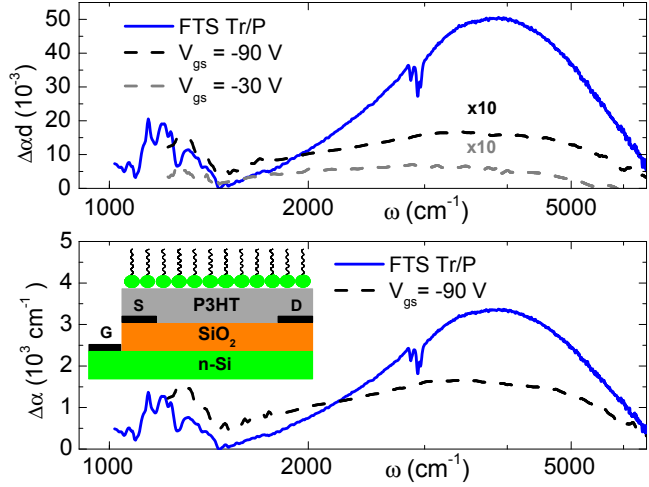}
     \caption{IR response of P3HT film modified by electrostatic doping in an OFET structure and under exposure to FTS fumes.  Top panel:  Mid-IR absorption $\Delta\alpha$d $=  (1- $T(V$_{gs})$/T(V$_{gs} = 0 V )$ for a device gated at V$_{gs}=-90 V$ (black dashed curve) and V$_{gs}=-30 V$ (gray dashed curve); $\Delta\alpha$d $=  (1-$T$_{tr}$/T$_p)$ for FTS-modification with no applied bias (blue solid curve).  Bottom panel:  Change in the absorption coefficient $\Delta\alpha$ for the same device.  Bottom inset:  Schematic of FTS-treated OFET device.} 
     \label{OFET}
\end{figure}

The three-terminal OFETs (Fig. \ref{OFET}) offer known advantages for studying electronic effects in organic semiconductors,\cite{bao-apl1996,zq-nanolett2006,zq-prb2007,bao-ofets,assadi-synthmet1990,brown-prb2001,friend-prl2003,dhoot-prl2006,braga-am2009,singh-armr2006,zq-prl2007,fischer-apl2007} and will serve as a reference system for characterizing the IR absorption associated with FTS effects on the electronic transport.  We first injected charges in a P3HT film by applying a gate voltage, forming a hole-accumulation layer in the polymer at the interface with SiO$_2$. With this gate-induced carrier density, we observed the spectroscopic fingerprints of electrostatic charge injection (black and grey dashed lines in top panel of Fig. \ref{OFET}).\cite{zq-nanolett2006} These include: (i) sharp peaks in the frequency range 1200 - 1400 cm$^{-1}$ attributable to infrared active vibrational modes (IRAVs) and (ii) a broad band centered around 3500 cm$^{-1}$ attributable to polaron absorption. The oscillator strength of these features, proportional to the area under $\Delta\alpha$d spectra, is voltage-dependent in agreement with published data.\cite{zq-nanolett2006}

After verifying the "fingerprint absorption" of the OFET, we subjected the same device to FTS fumes and repeated the measurements after 6 hours of exposure (blue line in top panel of Fig. \ref{OFET}).  In both cases, we observed sharp resonances in the frequency range between 1000 - 1500 cm$^{-1}$ and an absorption band centered at 4000 cm$^{-1}$.  The low frequency structure is reminiscent of IRAVs seen in the OFET. The apparent agreement is improved further with OFETs employing TiO$_2$ as the gate insulator, as SiO$_2$ allows observation of only the higher energy IRAVs.\cite{zq-nanolett2006}

\begin{figure}
\centering
   \includegraphics[width=3.375in, height=2.53in]{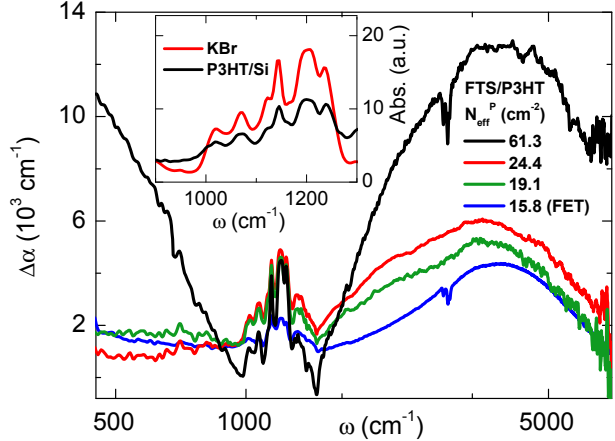}
     \caption{IR absorption of four different P3HT-based structures extended down to 40 cm$^{-1}$ to include the far-IR range of the spectrum.  The blue curve is FTS-induced absorption of the FET device presented in Figure \ref{OFET}.  All other curves are 2-probe (ungated) P3HT/Si structures doped with FTS.  The curves are labeled according to their integrated polaron spectral weight (detailed in text) and listed in order of polaron absorption strength.  Inset:  Vibrational spectrum of an FTS-treated P3HT/Si sample and FTS-coated KBr substrate.} 
     \label{IRdata}
\end{figure}

Upon detailed investigation, however, we found that the FTS layer itself grown on various substrates (including KBr and SiO$_2$) exhibits absorption resonances in the same frequency range as the electrostatically-induced IRAVs in P3HT (inset in Fig. \ref{IRdata}).  Here we compare the absorption of P3HT two-terminal device subjected to FTS fumes to that of FTS-coated KBr.  Note that clean KBr substrates are transparent in the mid-IR, and since FTS does not induce any measureable DC conductivity in KBr, we conclude that these peaks are primarily formed by the intrinsic vibrational modes of an FTS self-assembled network.  The two largest peaks are most likely due to the six CF$_2$ groups in each molecule, which typically have stretching frequencies in this range.\cite{lenk-langmuir1994,taso-langmuir1997,sinapi-tsf2007}

The broad absorption band at higher frequencies (centered at ~4000 cm$^{-1}$) in the IR spectra of both the gated P3HT devices and P3HT altered by FTS in Figure \ref{OFET} arises from optical transitions to energy levels within the energy gap. In nondegenerate ground-state polymers, excess charges lead to the formation of either polaron or bipolaron states within the gap.  This absorption from the localized polaron states is a characteristic spectral feature of charge injection in polymers.  The distinction between polarons and bipolarons in the context of FTS-doping will be discussed in the next section, and the broad absorption will be referred to as a \textquotedblleft polaron band\textquotedblright for simplicity. Similar polaron bands are also observed in photoexcited and chemically-doped polymers, and their line shape can be quantitatively described by models of polaron absorption.\cite{jiang-afm2002,vardeny-prl2002,heeger-rmp1988,heeger-prb1989,horovitz-ssc1982,vardeny-synthmet2004,vardeny-science2000} The appearance of the prominent polaron band in the FTS-induced absorption in the OFET provides evidence of oxidation or p-type doping in the polymer host.  In control experiments, we exposed KBr, Si, and GaAs substrates to FTS vapors. We did not find any significant modification of the absorption in the range where FTS-altered P3HT displays a polaron band.  Although the polaron bands and IRAV resonances appear simultaneously in IR spectra for electrostatically and/or chemically doped polymers,\cite{zq-nanolett2006,zq-prb2007,heeger-prb1988,heeger-rmp1988,heeger-prb1989,vardeny-science2000} the contributions from direct absorption of the FTS network discussed above do not allow unambiguous identification of IRAV features of FTS-controlled P3HT. 

We now analyze the strength of absorption of the gate-induced and FTS-induced polaron band in the P3HT film, both produced in the same OFET device.  It is customary to define an effective spectral weight 

\begin{equation}
N_{eff}^P = \int_{Pol}{(\Delta\alpha}d)d\omega,
\label{neffeqn}
\end{equation}

where the integration is done over the polaron band.  According to the oscillator strength sum rule,\cite{dressel} this spectral weight is proportional to the density of charges participating in an absorption feature: a polaron band in this case.  The $\Delta\alpha$d spectra plotted in Figure \ref{OFET} reveal that FTS-induced absorption is an order of magnitude stronger compared to that of the OFET device under V$_{gs} = -90 V$. This enhancement is at least in part due to the bulk nature of the doping in the P3HT film caused by the reaction with FTS. Doping throughout the thickness of the film results in a larger net absorption than that obtained from gating the same OFET (gate-induced charges are confined within a 1 - 2 nm thick accumulation layer).  The bottom panel in Figure \ref{OFET} compares only the change in the absorption coefficient, $\Delta\alpha$.  After properly correcting for the difference in thicknesses of the respective layers, it is apparent that the FTS reaction still yields higher doping levels than obtained by applying the highest sustainable gate voltage (blue and black curves in lower panel of Fig. \ref{OFET}).

Figure \ref{IRdata} shows IR absorption spectra for three different P3HT/Si two terminal structures and a P3HT OFET obtained in separate FTS treatments. The polaron band is reproduced in all spectra, reaffirming the assertion that this spectroscopic feature arises from doping by reaction with FTS. The oscillator strength of the polaron feature varies for different FTS treatment experiments, and can even vary across the same sample. The energy of the polaron absorption reflects the degree of order between polymer chains\cite{vardeny-science2000} and therefore can be expected to show some sample-to-sample variation. As shown in Figure \ref{IRdata}, we do find minor nonsystematic variation of the frequency of the polaron peak. Figure \ref{variation} shows data obtained by probing several different spots (separated by 1 mm) on the same two-terminal P3HT/Si sample. 1 mm is an enormous length scale compared to that of the features seen in AFM images of FTS-treated films,\cite{kao-afm2009} calling for further studies with nano-scale spatial resolution that are now possible.        
 
Structures which showed the most intense polaron band oscillator strengths reveal an additional feature in the far-IR absorption (black trace in Fig. \ref{IRdata}, Fig. \ref{variation}).  The monotonic increase of this far-IR absorption towards lower frequencies (for $\omega <$ 1000 cm$^{-1}$) is the Drude-like response of delocalized charge carriers,\cite{dressel} and apparently signifies metallic transport in FTS-doped films.  The low-energy absorption persists at low temperatures, consistent with the notion of metallic transport, with a modified $\omega$-dependence to be reported in a separate publication. From the data in Fig. \ref{variation}, we extract a two-dimensional (2D) conductivity of $\sigma_{DC}^{2D} = $.970 $k\Omega^{-1}$, which is consistent with the highest values seen in transport. \cite{kao-afm2009} The coexistence of metallic (delocalized) and polaronic (localized) states may point to a phase-separated system.   

Additionally from Fig. \ref{variation}, the Drude absorption appears to be more uniform over the large length scale considered. This suggests that the variation in the mid-IR absorption is likely due to light scattering, as opposed to inhomogeneities in the FTS treatment. Some inevitable swelling occurs in doped films, altering the surface morphology and leading to a reduced IR signal through regions with large surface irregularities. IR microscopy at the nano-scale is necessary to resolve such issues of inhomogeneity.\cite{mumtazPRB2009}     

\section{Analysis and Discussion}

\begin{figure}
\centering
   \includegraphics[width=3.375in, height=2.53in]{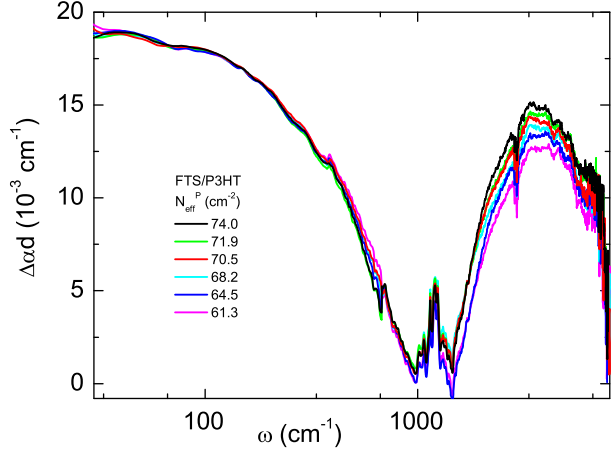}
     \caption{IR absorption spectra collected from different spots on a single $1x1$ cm$^{2}$ P3HT/Si sample doped with FTS to saturation. Probed locations are separated by 1 mm} 
     \label{variation}
\end{figure} 
 
\begin{figure}
\centering
   \includegraphics[width=3.375in, height=2.53in]{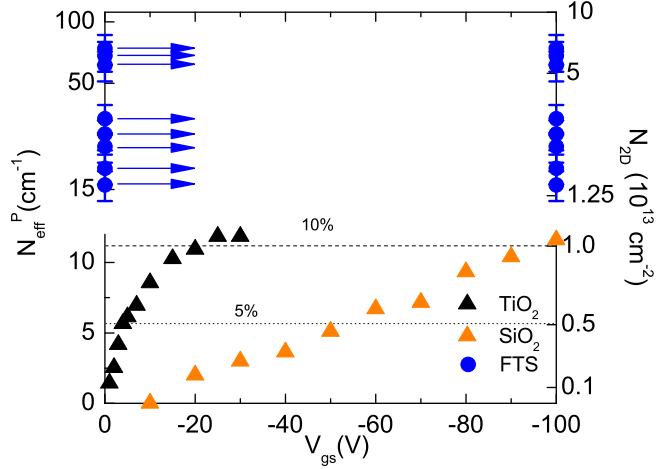}
     \caption{Effective 2D spectral weight of the polaron band $N_{eff}^P$ and corresponding 2D carrier density plotted as a function of the applied gate voltage for SiO$_2$ ($\mid$V$_{gs}$$\mid$=0-100V) and TiO$_2$ ($\mid$V$_{gs}$$\mid$=0-30V)-based FET devices (triangles).  The blue circles are $N_{eff}^P$ and $N_{2D}$ associated with the FTS-doped P3HT structures, with error bars due to the uncertainties in film thickness. Dotted and dashed lines indicate doping levels of 5\% and 10\%, respectively, for P3HT.} 
     \label{neff}
\end{figure}

The gate-induced or doping-induced carrier density in P3HT can be extracted from the spectroscopic data.  Note that an OFET can be modeled as a parallel plate capacitor in which the charge density $N_{2D}$ is proportional to the applied gate voltage: 

\begin{equation}
N_{2D} = \frac{\kappa \epsilon_{0}}{eL} V_{gs},
\label{n2d}
\end{equation}

where $\kappa$ and $L$ are the dielectric constant and thickness of the gate insulator, respectively.  Naturally, $N_{2D}$ is the same at both interfaces of the dielectric (i.e., in the gate electrode and in the OFET accumulation layer).  The 2D carrier density in an accumulation layer in the P3HT-based OFET can thus be obtained directly from the applied gate voltage.  Using the effective spectral weight $N_{eff}^P$ defined earlier (Eqn. \ref{neffeqn}), there exists a one-to-one correspondance between the carrier density and the strength of the polaron absorption. The validity of this approach is confirmed by the linear dependence of $N_{eff}^P$ on V$_{gs}$ observed in OFETs with SiO$_2$ insulator (red triangles in Fig. \ref{neff}).  Previous work on SiO$_2$-based P3HT OFETs has shown a 2D carrier density reaching 10$^{13}$ holes/cm$^2$ at V$_{gs} = -100 V$ in structures with similar thickness of the gate insulator\cite{zq-prb2007} and is limited by the break-down of the insulator.

The doping-induced carrier density in FTS-treated P3HT can then be obtained by integrating the absorption spectra: 

\begin{equation}
N^P_{eff, FTS} = \int_{Pol}{(\Delta\alpha)d\omega}.
\label{nefffts}
\end{equation}

Figure \ref{neff} shows a comparison of  for the OFET- and FTS-based methods of introducing charge carriers.  It is evident from the plot that much higher carrier densities are attainable in FTS-doped P3HT, delivering as much as an order of magnitude enhancement over OFETs.

The capacitor model indicates that carrier densities in OFETs can in principle be increased with an appropriate choice of gate insulator.  However, the experimental reality of OFETs employing high dielectric constant oxides as gate insulators is more complex than this simple conjecture.  In Figure \ref{neff} we plot data for OFETs with TiO$_2$ ($\kappa=37$) and SiO$_2$ ($\kappa=3.9$),\cite{zq-nanolett2006} and indeed devices employing TiO$_2$ dielectrics show strong enhancement of the carrier density (blue triangles), but only at small V$_{gs}$ in the regime where the carrier density still varies linearly with the gate voltage.  At V$_{gs}$ approaching $-30 V$ ideal OFET behavior is arrested due to high leakage current preempting breakdown at yet higher V$_{gs}$.  In our experience, TiO$_2$ dielectrics yield no sizable improvement to the maximum possible carrier densities in OFETs compared to SiO$_2$-based devices.  The data in Figure \ref{neff} reaffirm that FTS modification of polymers offers a viable means to achieve carrier densities significantly exceeding that of oxide-based OFETs.  It is important to note that the FTS-induced carrier densities shown in Fig. \ref{neff} are calculated considering only the polaron band. Including the low-energy Drude absorption in the integral in Eqn. (\ref{nefffts}) puts the density of injected carriers from FTS molecules slightly higher in the 10$^{14}$ holes/cm$^2$ range for the highly-doped films.

\begin{figure}
\centering
   \includegraphics[width=3.375in, height=2.53in]{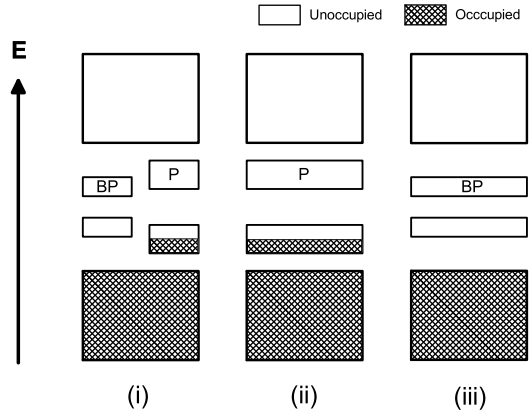}
     \caption{Schematic energy level diagram displaying three possible scenarios for an insulator-metal transition in P3HT upon doping: (i) a first-order transition from a bipolaron to a polaron lattice, leading to a partially-filled band; the (ii) polaron or (iii) bipolaron band broadens to merge with the valence band.}
     \label{MIT}
\end{figure}

Data reported in Figs. \ref{IRdata} - \ref{neff} may reveal additional details about the insulator-metal transition that occurs in doped P3HT. In Ref. \onlinecite{zq-prb2007}, three possible theoretical scenarios were suggested to describe the IMT in P3HT: (i) a first-order transition between a bipolaron and a polaron lattice, (ii) a gap closure between the polaron band and the valence band, and (iii) a gap closure between the bipolaron band and the valence band. A schematic diagram of these three possibilities for a transition is displayed in Fig. \ref{MIT}. The critical doping concentration for both scenarios (i) and (iii) was predicted to be near 10\%, or \~ 10$^{13}$ /cm$^2$, while for scenario (ii) it is closer to 5\%.  It is evident from Fig. \ref{neff} that 10$^{13}$ /cm$^2$ is barely attainable in gated OFETs. Despite being on the cusp of the IMT, no signs of metallic conductivity are evident in the OFET absorption data at the highest gate voltages. Using a theoretical model originally proposed by Brazovskii and Kirova,\cite{BKjetp1981} the lowest optical transition frequency was calculated to be 0.16 eV for polarons, and 0.45 eV for bipolarons.\cite{zq-prb2007} With these energies in mind, P3HT OFET data show one broad absorption at 3500 cm$^{-1}$ (0.44 eV) at all dopings up to 10\%, indicating that this absorption is likely due to bipolarons.  This suggests a merging of the valence and bipolaron band as the mechanism by which an IMT occurs.   

Here, we have increased the carrier density in P3HT an order of magnitude further, with clear signs of metallicity, and both the peak structure and position of the mid-IR absorption band have not changed appreciably.  Therefore, in light of the three IMT scenarios presented earlier, the new data would provide additional evidence for the physical picture of a bipolaron band merging with the valence band to form a partially-filled (metallic) energy band. Previous work on P3HT has led to numerous interpretations of the low-energy excitations in the absorption spectrum for the various methods of introducing excess charges (electrostatic, chemical, etc.).\cite{zq-prb2007,jiang-afm2002,vardeny-science2000,sirringhausNature1999, wohlgenanntPRB2004}  To evaluate these interpretations as well as the models used to calculate polaron and bipolaron energies, effects such as disorder must be taken into account.  This is especially true in the case of FTS doping, where the penetration of FTS molecules into the bulk of the P3HT film may have a significant effect on the local order of the polymer chains. Perhaps more importantly, most of the theoretical models used to calculate these energies assume non-interacting polarons.  In order to completely capture all of the essential physics at such high carrier densities near the IMT, interactions between the polarons must be taken into consideration.\cite{iadonisiEuro1999,fratiniPRB2009}
 	
The previous interpretation of a continuous transition from an insulating to metallic state is based upon the assumption of a homogeneous film.  This may not necessarily be the case.  The coexistence of metallic and localized features in the IR absorption data (Figs. \ref{IRdata} and \ref{variation}) for FTS-doped films is indicative of phase separation.  Earlier, Dhoot et al. estimated that a small fraction ($\sim$ 1\%) of the field-induced carriers in regio-regular P3HT OFETs occupy delocalized states at high carrier densities, while most charges remain in the lower-energy localized states.\cite{dhoot-prl2006} Phase separation may be a common attribute among conducting polymers.  Earlier work on heavily-doped polyacetylene has demonstrated metallic absorption appearing simultaneously with localized excitations (IRAVs). Recent work on polyaniline, however, has demonstrated a metallic response in reflectance simultaneously with reduced-strength IRAVs.\cite{heeger-nature2006} The strong absorption of FTS molecules obscures the IRAV spectrum in FTS-doped P3HT, but it is important to note that the metallic response observed in Fig. \ref{IRdata} occurs simultaneously with the highest polaron absorption, relative to other treated samples. Truly resolving these issues certainly requires these effects to be thoroughly studied at the nano-scale.   

\section{Conclusions}

The IR data (Figs. \ref{OFET}, \ref{IRdata}) confirm the assertion that the large increase in DC conductivity in FTS-treated structures is based on doping via electron transfer from the conjugated polymer to electronegative FTS molecules located nearby the polymer chains.  The data in Figure \ref{neff} demonstrate that carrier densities in FTS-doped films are significantly higher than those in OFETs, with the highest values reaching nearly 10$^{14}$ holes/cm$^2$.  Such strong doping is indicated by the significantly enhanced polaron band and the strong far-infrared Drude-like absorption. The range of carrier densities is consistent with the variation seen in transport measurements of FTS-treated P3HT, which typically have saturated DC conductivities in the range $\sigma_{DC}^{2D}$ = 10$^{-4}$ - 10$^{-3}$ $\Omega^{-1}$.\cite{kao-afm2009}  The Drude-like form of the far-IR absorption is characteristic of a free carrier response.  These data therefore indicate that FTS-controlled films can be doped to sufficiently high carrier densities (of order 10$^{14}$ holes/cm$^{2}$ or 10$^{21}$/cm$^{3}$) to be on the metallic side of the insulator-to-metal transition.  The transition to the metallic state in this highly-doped regime is in accord with the analysis of required carrier densities.\cite{zq-prb2007} With the number of carriers surpassing 10$^{14}$ holes/cm$^2$, FTS-doping provides an opportunity to study the nature of the insulator-to-metal transition in conjugated polythiophenes. When combined with IR nanoscopy, these studies can directly address competing physical descriptions of the IMT, such as phase separation or a continuous transition to metallic states.

\section*{Acknowledgments}
This work is supported by AFOSR Grant No. FA9550-09-1-0566 and University of California.

\end{document}